# Two-stage Respiratory Motion-resolved Radial MR Image Reconstruction Using an Interpretable Deep Unrolled Network

Shanshan Shan[1], Hongli Chen[2], Yuhan Wei[3], Peng Wu[4], Yang Gao[5], Tess Reynolds[6], Paul Liu[6], Jialiang Zhang[1], Qidi Luo[1], Chunyi Liu[7], Paul Keall[6], Feng Liu[2], Yaqin Zhang[3*], David E. J. Waddington[6], Mingyuan Gao[8*]

[1] State Key Laboratory of Radiation Medicine and Protection, School of Radiation Medicine and Protection, Collaborative Innovation Center of Radiological Medicine of Jiangsu Higher Education Institutions, Soochow University, Suzhou 215123, China.

[2] School of Electrical Engineering and Computer Science, The University of Queensland, Brisbane, QLD 4072, Australia.

[3] Department of Radiology, The Fifth Affiliated Hospital of Sun Yat-sen University, Zhuhai 519000, China.

[4] Philips Healthcare, Shanghai, China.

[5] School of Computer Science and Engineering, Central South University, Changsha 410083, China.

[6] Image X Institute, Sydney School of Health Sciences, Faculty of Medicine and Health, University of Sydney, Sydney, NSW 2015, Australia.

[7] Medical School, Nanjing University, Nanjing 210093, China.

[8] School of Life Sciences, Soochow University, Suzhou 215123, China.

\* **Corresponding Author**: Mingyuan Gao (gaomy@suda.edu.cn) and Yaqin Zhang (zhyaqin@mail.sysu.edu.cn)

# Abstract

Due to the prolonged MRI encoding process, respiratory motion can cause undesired artifacts and image blurring, degrading image quality and limiting clinical applications in abdominal and pulmonary imaging. In this work, we develop a two-stage respiratory motion-resolved radial MR image reconstruction pipeline using an interpretable deep unrolled network (MoraNet), enabling high-quality imaging under free-breathing conditions. Firstly, low-resolution images are reconstructed from the central region of successive golden-angle radial k-space to extract respiratory motion signals. The binned k-space data based on the respiratory signal are then used to reconstruct the motion-resolved high-resolution image for each motion state. The MoraNet applies nonuniform fast Fourier transform (NUFFT) to operate radial encoding and convolutional neural network (CNN) modules to conduct image regularizations. The MoraNet was trained on retrospectively acquired lung MRI images for both fully sampled and undersampled acquisitions. The performance of the proposed method was evaluated on digital CT/MRI breathing XCAT (CoMBAT) phantom data, QUASAR motion phantom data acquired from a 1.0T MRI scanner and volunteer chest data acquired from a 1.5T MRI scanner. The MoraNet pipeline was compared with motion-averaged reconstruction and a conventional compressed sensing (CS)-based method in terms of structural similarity (SSIM), root mean square error (RMSE) and computation time. Simulation and experimental results demonstrated that the proposed network could provide accurate respiratory signal estimation and enable effective motion correction. Compared with the CS method, the MoraNet preserved better structural details with lower RMSE and higher SSIM values at acceleration factor of 4, and meanwhile took ten-fold faster inference time. The MoraNet can achieve fast, dynamic, motion-resolved image reconstructions and thus has the potential to facilitate clinical translations.

# 1. Introduction

Liver and lung cancers are the leading causes of death in the world, accounting for over 20% of total deaths [1]. Magnetic resonance imaging (MRI) technique has been increasingly applied for the early-stage cancer diagnoses due to the superior soft-tissue contrast and non-radiation advantage [2]. However, the respiratory motion can cause undesired artifacts and degrade image quality in abdominal and pulmonary MRI because of the prolonged encoding process [3]. During a single breathing cycle, the position of the abdominal organs can change by several centimeters [4], which poses a significant challenge for accurate abdominal MR imaging. Typically, MR images at different respiratory states are essential to analyze chest volumetric changes and pulmonary functions [5]. Whereas respiratory-induced blurring and artifacts can lead to image misregistration and potentially inaccurate pulmonary function evaluation.

Recently, golden-angle radial sequences have been widely developed to provide free-breathing and motion-robust MRI reconstructions. Unlike traditional Cartesian sequences, the k-space center of radial acquisitions is repeatedly sampled and thus can be used to estimate the respiratory motion signals without requiring additional sensors, serving as a self-navigator [6]. The radial k-space center (direct current, DC) data represent the average signal intensity of the whole excitation volume, which changes with the respiratory motion and thus can be used as a self-navigation signal [7-10]. However, DC signals are sensitive to some acquisition factors, such as the rotating of readout direction, gradient induced heating and bulk motion, often leading to the image reconstruction inaccuracy [11, 12]. Alternatively, image-based self-navigated approach exploits low-resolution dynamic images to directly measure respiratory motion and this strategy has significantly improved the reconstruction accuracy in contrast to the DC-based method [13, 14]. However, previous studies have not investigated and reported evaluations on the accuracy of the respiratory signal estimation, as the ground truth respiratory signal is often not available.

Based on the respiratory signal, the acquired k-space data are normally binned into multiple respiratory motion states and motion-resolved images can be reconstructed by compressed sensing (CS) algorithms. For instance, XD-GRASP deploys the total variation (TV) along the respiratory-state dimension as the regularization term to impose temporal sparsity, and reconstructs the motion-resolved images from the undersampled binned k-space for each respiratory phase [15, 16]. Motion fields and low-rank constraints have also been integrated in the CS-based methods to provide free-breathing and motion-compensated pulmonary MRI reconstruction [17-19]. However, it is often empirical and tedious to select the regularization weighting parameters in the CS-based methods [20, 21]. In addition, the iterative process is computationally expensive and therefore is impractical for clinical implementations.

Deep neural networks have been increasingly investigated for solving inverse non-Cartesian MRI reconstruction problems and removing respiratory motion artefacts [22, 23]. The image-domain based methods such as Phase2Phase [24] and XD-Net [25] have been proposed to learn the mapping from undersampled/aliased images to the motion-resolved/unaliased images, showing improved dynamic liver imaging performance.

Deep unrolled networks (e.g., stDLNN [26], DCReconNet [27] and RebinNet [28]) have been designed to incorporate MR physics in the network architecture to perform non-uniform MRI image reconstructions, exhibiting better interpretability and generalization ability than the image-domain based networks. However, the deep unrolled network has not been applied in the dynamic low-resolution image reconstruction for the respiratory motion estimation. In addition, the radial trajectories of binned k-space data are randomized, and the reconstruction performance of deep unrolled network on binned k-space data has not been investigated and reported to the best of our knowledge. In this work, we develop and investigate a two-stage free-breathing respiratory motion-resolved radial MR image reconstruction (referred to as MoraNet) pipeline for motion signal estimation and motion artifacts correction. The proposed pipeline leverages the advance of interpretable deep unrolled network architectures for both low-resolution and high-resolution image reconstructions at two stages. Firstly, the central region of successive radial k-space data is fed into the MoraNet to reconstruct dynamic low-resolution images, which are used to measure the respiratory motion signals. The whole k-space data are then binned for each motion state based on the measured motion signals and the high-resolution motion-resolved images are reconstructed by the MoraNet pipeline. The proposed network consists of CNN modules as regularization terms and nonuniform fast Fourier transform (NUFFT) as radial encoding operator. The MoraNet was trained on retrospectively acquired images from lung cancer patients and then evaluated on digital CT/MRI breathing XCAT (CoMBAT) phantom data, prospective QUASAR motion phantom data from a 1.0T MRI scanner, and volunteer pulmonary data from a 1.5T MRI scanner, respectively. The estimated respiratory/motion signals at stage 1 from CoMBAT and QUASAR phantoms were compared quantitatively with the ground truth (GT). The MoraNet-reconstructed images at stage 2 were compared with the motion-averaged and conventional CS reconstructed images in terms of the image quality and computation time for fully sampled and subsampled acquisitions.

## 2. Methods and Materials

### 2.1 MoraNet reconstruction workflow

The MoraNet is developed for reconstructing respiratory motion-resolved images from successively acquired golden-angle radial k-space data with free-breathing. As shown in Figure 1(a), the central region of radial k-space data is initially used to reconstruct the dynamic low-resolution images. The respiratory motion signal is estimated from the low-resolution images and the whole k-space data are retrospectively binned into n (n = 10) motion states at stage 1. Afterwards, the binned k-space data for each motion state is fed into the MoraNet to reconstruct high-resolution motion-resolved images at stage 2, as shown in Figure 1(b). The MoraNet architecture consists of seven iterative soft shrinkage-thresholding layers and each layer starts with a data fidelity module, followed by a CNN-based regularization term (Figure 1(c)). The NUFFT [20] algorithm is applied for the radial encoding operation in the data fidelity module. The CNN-based regularization term includes nonlinear forward and backward transforms with a soft thresholding operation, which is designed to remove the image artifacts induced by the k-space undersampling. Each nonlinear transform combines a rectified linear unit (ReLU) and two convolutional operators. A skip connection with a residual block is applied to further facilitate the network training performance.

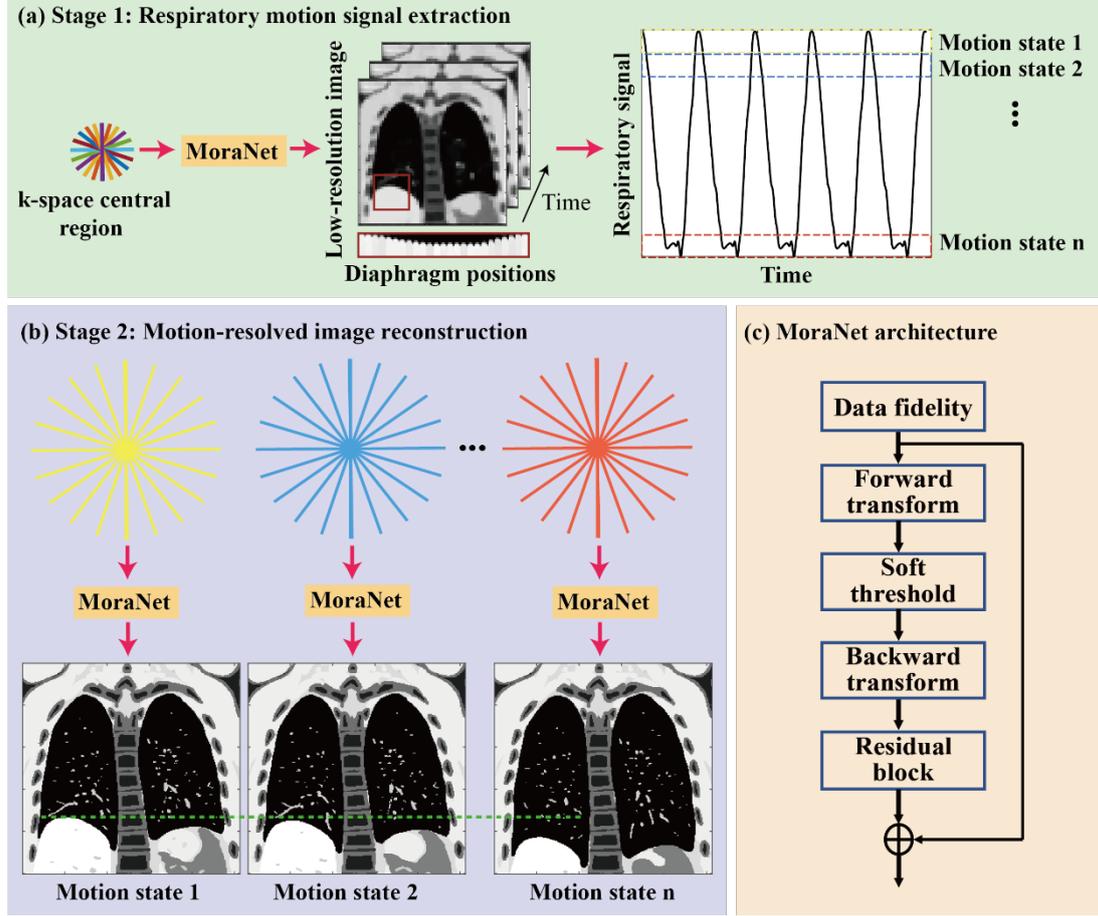

Figure 1 The overall workflow of MoraNet reconstruction. (a) At stage 1, dynamic low-resolution images are reconstructed to extract respiratory motion signals. (b) At stage 2, the whole k-space data are grouped into *n* bins and the high-resolution motion-resolved images for each motion state are reconstructed. (c) MoraNet network architecture.

### 2.1.1 Dynamic low-resolution image reconstruction and respiratory motion signal extraction

Given the successively acquired golden-angle radial k-space data, the dynamic low-resolution image $\tilde{x}_{lr}$ reconstructed by the MoraNet at stage 1 can be formulated as:

$$\tilde{x}_{lr} = \underset{\tilde{x}_{lr}}{\operatorname{argmin}}\{\|F\tilde{x}_{lr} - mb_i\|_2^2 + CNN(\tilde{x}_{lr})\} \qquad (1)$$

where $b_i$ represents $i$ ($i = 32$) successively acquired radial k-space spokes and $m$ denotes the mask used to select the central region of k-space measurements. $F$ is the radial Fourier transform operator, which can be implemented by the type-II NUFFT algorithm [20]. $\|F\tilde{x}_{lr} - mb_i\|_2^2$ is the data fidelity term to minimize the difference between estimated and measured central region of radial k-space. $CNN(\tilde{x}_{lr})$ denotes the CNN-based latent regularization to reduce artifacts and improve image quality.

Based on a series of dynamic low-resolution images, the 1D surrogate signal of respiratory motions (e.g., diaphragm positions in Figure 1(a)) can be derived and used for the subsequent k-space data binning operation.

### 2.1.2 K-space binning and motion-resolved image reconstruction

According to the measured respiratory motion signal, the respiratory circle can be separated into several different motion states and the whole k-space radial data are binned together for each motion state, as shown in Figure 1(b). The motion-resolved image for each motion state is then reconstructed by MoraNet at stage 2, governed by the equation below:

$$x_j = \underset{x_j}{\operatorname{argmin}} \left\{ \left\| F x_j - b_j \right\|_2^2 + CNN(x_j) \right\} \qquad (2)$$

where $x_j$ is the motion-resolved image for $j_{th}$ motion state and $b_j$ is the corresponding binned k-space data. It is noted that the radial trajectory of the binned k-space data $b_j$ has been randomized. $F$ represents the NUFFT operator and $CNN(x_j)$ is the regularization term. The first term $\left\| F x_j - b_j \right\|_2^2$ enforces the data fidelity and the second term $CNN(x_j)$ promotes image sparsity.

## 2.2 Training data preparation

3000 retrospectively acquired lung MRI images from cancer patients were augmented five times through rotation and flip operations, and a total of 15000 lung images were used to train the MoraNet. The lung MRI data were split in the ratio of 10:1 for training and testing. The acquisition parameters of the lung cancer dataset are detailed in Refs [29, 30]. For stage 1, the lung data were downsampled to low-resolution images of size 64×64, which were used as label images during the model training. The low-resolution images were then encoded with successive tiny golden-angle (20.89°) and golden-angle (111.25°) radial trajectories to generate the k-space data using NUFFT operations. Each readout spoke had two-time oversampled (128) data points. Different k-space spokes (64, 32, 24 and 16) were generated and fed into the network as the input, respectively. For stage 2, the high-resolution lung images were used as GT for the network training. Radial spokes randomly selected from successive tiny golden-angle and golden-angle radial trajectories were used to simulate the fully-sampled k-space data for each lung image. The generated k-space data were also undersampled with acceleration factors (AF) of 2 and 4. The fully-sampled and undersampled k-space data were used as the input for the MoraNet training process, respectively.

## 2.3 Testing data acquisition

Free-breathing lung cancer MRI data were simulated from the digital CoMBAT phantom [31] with balanced steady-state free precession (bSSFP) sequence, TR/TE = 10/4.5 ms, image resolution = 256×256, total spoke number = 4908, readout samples = 512. A QUASAR motion phantom [32, 33] was scanned from a 1T MRI-Linac [34] system to acquire successive tiny golden-angle (20.89°) radial k-space data. The acquisition parameters were: GRE sequence, channel number = 8, TR/TE=10/5 ms, image resolution = 128×128, total spoke number = 10050, and readout samples = 256.

Six free-breathing volunteers were scanned with a multi-channel torso coil in a clinical 1.5T Ambition Philips scanner using golden-angle (111.25°) acquisitions and the imaging parameters were: GRE sequence, TR/TE = 8/2.3 ms, image resolution = 448×448, total spoke number = 25000, and readout samples = 600. The binned volunteer radial k-space data for each motion state at stage 2 were retrospectively and prospectively undersampled by AFs of 2 and 4 to evaluate the performance of the presented workflow.

## 2.4 Model training and evaluation methods

The MoraNet was trained on a high-performance workstation utilizing an Nvidia Tesla V100 P32 graphical processing unit (GPU). 100 epochs with ~20 h, a batch size of 32 and Adam optimizer [35] are utilized during the training process. The learning rate was 0.001 and 0.0001 for the first half and the remaining epochs, respectively. The loss function was calculated by the mean square error (MSE) for the network training. All human studies were conducted with the approval of the Institutional Review Board (IRB).

In this work, the motion-averaged images were reconstructed by the conventional NUFFT algorithm and compared with motion-resolved images (including end-expiration, intermediate and end-inspiration motion states). The CS-based iterative algorithm and the proposed MoraNet were also implemented to reconstruct undersampled multi-channel radial k-space data (AFs = 2 and 4), and the MoraNet-reconstructed images with fully sampled acquisitions (referred to as MoraNet-FS) were served as reference. The k-space data from each channel were used as inputs and the sum-of-squares (SoS) [36] was performed on all single-channel reconstructed images to obtain the coil-sensitivity-combined images. To quantitatively evaluate image quality, the structural similarity index (SSIM) and root mean square error (RMSE) were calculated using the reconstructed and reference images.

## 3. Results

### 3.1 Motion signal extraction

The simulated CoMBAT phantom (Figure 2 (a)) and experimental QUASAR motion phantom (Figure 2 (e)) images were used to evaluate the accuracy of the motion signal extraction at stage 1 in the MoraNet pipeline. The respiratory/motion signals estimated from reconstructed low-resolution CoMBAT and QUASAR phantom images with different k-space spokes were compared with the ground truth in Figure 2 (b-d) and Figure 2 (f-h). Good consistency is shown between estimated (orange line) and GT (blue line) motion signals in Figure 2 (b) and Figure 2 (c), where 64 and 32 spokes were used for image reconstruction, respectively. Whereas, discrepancy between estimated and GT respiratory signal is displaced in Figure 2 (d) with 16-spoke reconstruction. Similarly, the extracted QUASAR motion signal in Figure 2 (f) (64 spokes) and Figure 2 (g) (32 spokes) has better accuracy than that in Figure 2 (h) (24 spokes). The quantitative analysis of motion signal displacements is shown in Figure 3. The displacement RMSE of the testing phantoms is within 0.5 mm and the median value is less than 2 mm for spokes1 and spokes2, suggesting accurate respiratory signal measurement at stage 1 of the MoraNet pipeline. In comparison, the median displacement for spokes3 is larger than 2 mm, showing degraded motion estimation accuracy, which is consistent with the results in Figure 2.

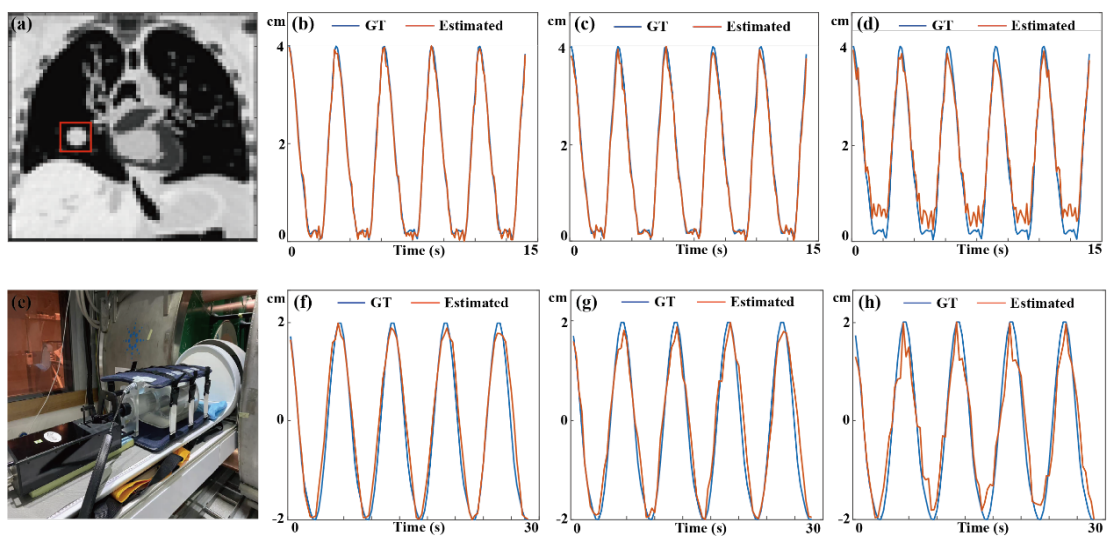

Figure 2 (a) CoMBAT phantom. Tumour positions (red rectangle) on simulated low-resolution CoMBAT phantom images were calculated to estimate respiratory signals. (b-c) Spokes1, Spokes2 and Spokes3 respiratory signal. The CoMBAT respiratory signal was extracted from low-resolution images reconstructed with 64 (b), 32 (c) and 16 (d) spokes, respectively. (e) QUASAR motion phantom experimental setup. (f-h) Spokes1, Spokes2 and Spokes3 motion signal. The QUASAR motion signal was extracted from images reconstructed with 64 (f), 32 (g) and 24 (h) spokes, respectively. The estimated CoMBAT and QUASAR motion signals from dynamic low-resolution images at stage 1 were compared with the GT.

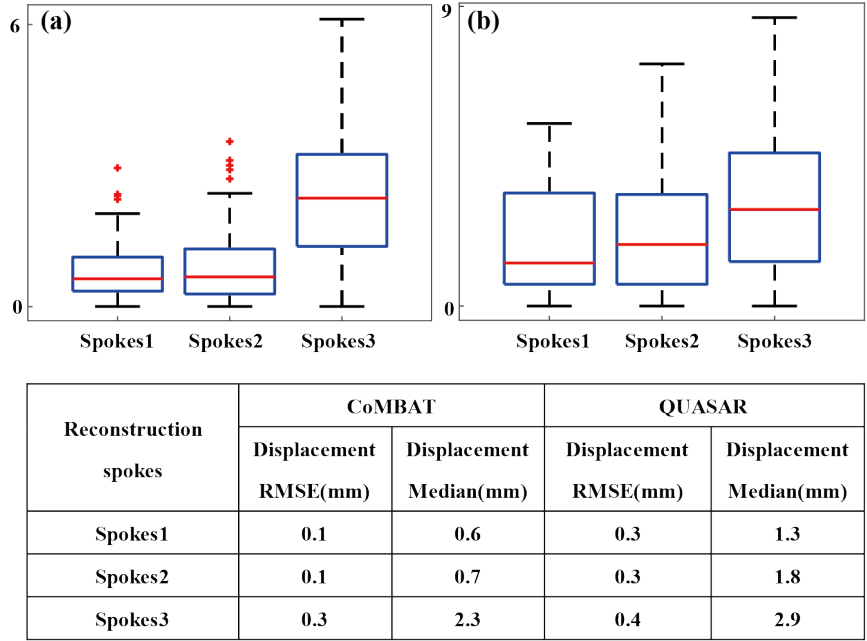

| Reconstruction spokes | CoMBAT | | QUASAR | |
|---|---|---|---|---|
| | Displacement RMSE(mm) | Displacement Median(mm) | Displacement RMSE(mm) | Displacement Median(mm) |
| Spokes1 | 0.1 | 0.6 | 0.3 | 1.3 |
| Spokes2 | 0.1 | 0.7 | 0.3 | 1.8 |
| Spokes3 | 0.3 | 2.3 | 0.4 | 2.9 |

Figure 3 (a) CoMBAT displacement (mm). (b) QUASAR displacement (mm). The boxplots of the motion signal displacements for CoMBAT (a) and QUASAR (b) phantoms are displayed. Quantitative results including the RMSE and Median value of motion displacements from reconstructed CoMBAT and QUASAR phantom images using different spokes are shown in the table.

## 3.2 CoMBAT phantom reconstruction

The performance of the MoraNet method in reconstructing digital CoMBAT phantom images is shown in Figure 4. Image blurring and motion artifacts are noticeable in the motion-averaged images reconstructed by the conventional NUFFT algorithm. Whereas, the MoraNet significantly reduced the motion blurring as shown in the end-expiration, intermediate and end-inspiration motion-resolved images. As indicated by the dashed line, the tumour position changed from top to bottom between the end-expiration and the end-inspiration state. The RMSE and SSIM values of 19 successive motion-averaged and motion-resolved images at the same timepoints were calculated and plotted in Figure 4. The motion-resolved images have considerably lower RMSE (0.17 median value) and higher SSIM levels (0.84 median value) than the motion-averaged ones (0.2 RMSE and 0.46 SSIM median values), indicating improved image quality.

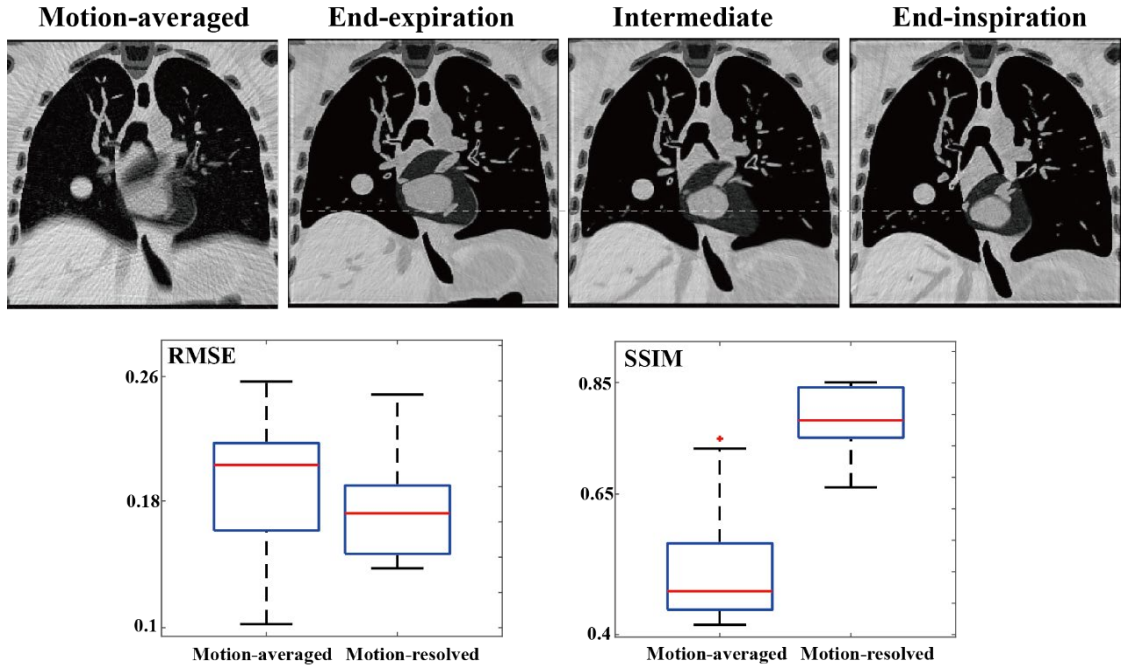

Figure 4 Motion-averaged and motion-resolved (including end-expiration, intermediate and end-inspiration motion states) digital CoMBAT phantom images reconstructed by the MoraNet pipeline. Boxplots of RMSE and SSIM values across 19 successive motion-averaged and motion-resolved images at the same time points were plotted. Minimal and maximal values with first quartile (25%), median (50%) and third quartile (75%) were statistically plotted, and red crosses represent outliers.

## 3.3 QUASAR motion phantom reconstruction

An MR compatible QUASAR motion phantom was scanned using a 1T MRI-Linac system with successive tiny golden-angle radial acquisitions. Compared with the motion-averaged images, less blurring and fewer artifacts are presented in the motion-resolved images as indicated by the red arrows in Figure 5. Line profiles along the yellow lines demonstrated that the MoraNet pipeline resulted in sharper edges than the conventional NUFFT reconstruction without the k-space binning operation. The dynamic high-resolution motion-resolved images in Figure S1 show sharper edges and less artifacts than the motion-averaged images.

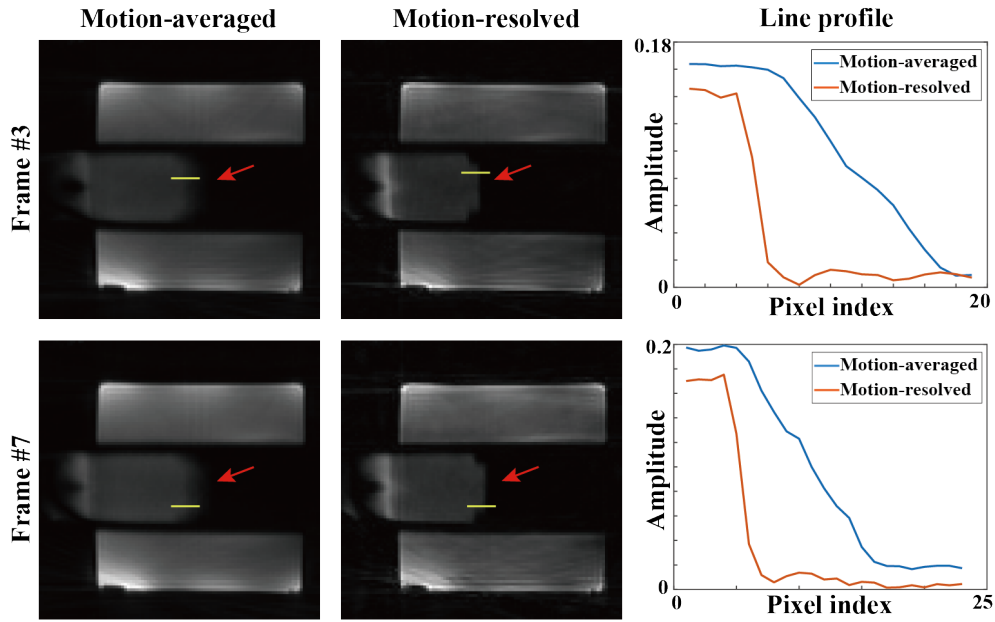

Figure 5 MoraNet reconstruction results based on the prospectively acquired motion phantom data. An MR compatible QUASAR motion phantom was scanned on a 1T MRI-Linac system. Motion-averaged and motion-resolved phantom images at frame #3 and frame #7, and line profiles along the yellow lines are shown.

### 3.4 In Vivo volunteer results

Fully sampled free-breathing volunteer chest images reconstructed by the conventional NUFFT method and the proposed MoraNet workflow are shown in Figure 6. Image blurring and artifacts are noticeable at the lung-liver interface (diaphragm) as indicated by the red arrows in the motion-averaged images. The diaphragm structure is sharper in the motion-resolved images and the diaphragm position clearly changes from the end-expiration state to the end-inspiration state, which is consistent with the results of Figure 4. The dynamic motion-resolved images in Figure S2 suggest that better image quality and clearer structural details are achieved by the MoraNet reconstruction pipeline.

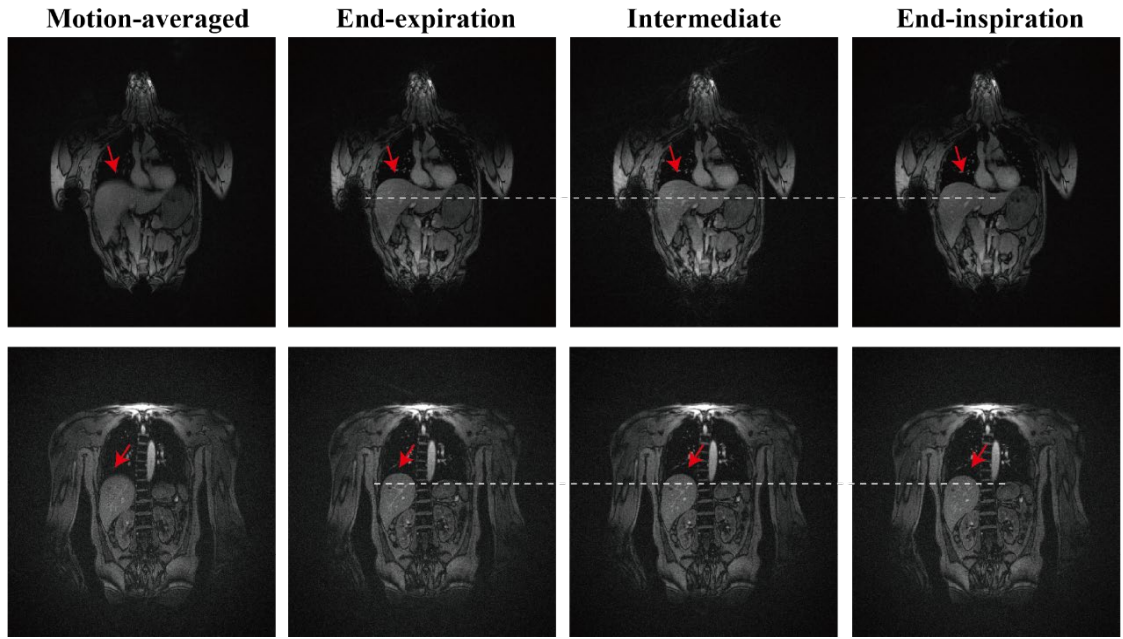

Figure 6 MoraNet reconstruction results on the fully sampled volunteer chest data processed with motion-averaged and motion-resolved (including end-expiration, intermediate and end-inspiration states) methods, respectively. Red arrows indicate the sharpness of the lung-liver interface.

The reconstruction performances of the conventional CS and the MoraNet methods on retrospectively undersampled binned k-space data (at stage 2) with AFs = 2 and 4 were compared in Figure 7. Undesired artifacts and image detail loss were observed on motion-averaged images, as pointed out by the red arrows. These motion artifacts and image blurring are significantly reduced on MoraNet-FS images, which are used as reference for the undersampling image reconstructions. For the low acceleration factor (AF = 2), the MoraNet achieved comparable reconstruction results with the conventional CS method for same RMSE and SSIM values. Whereas, the MoraNet-reconstructed image shows finer structural details with lower RMSE and higher SSIM values than the CS-reconstructed image at AF = 4, demonstrating better reconstruction performance.

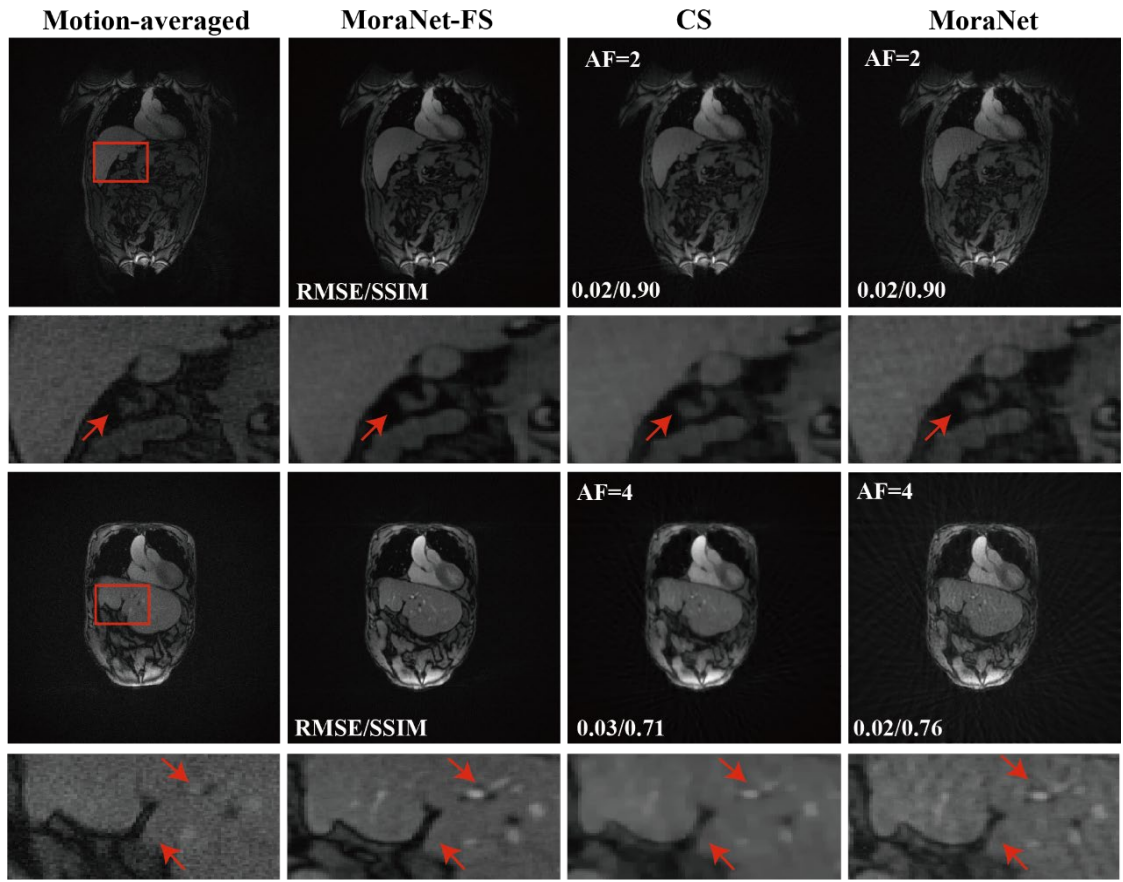

Figure 7 MoraNet reconstruction performance on the fully sampled (motion-averaged and MoraNet-FS) and the retrospectively undersampled (CS and MoraNet) binned k-space data at AFs = 2 and 4, respectively. Zoomed regions (red rectangle) are shown at the bottom of each reconstructed image and red arrows indicate the image structural details.

The MoraNet-reconstructed images on the prospectively undersampled binned k-space data (AFs = 2 and 4, respectively) at stage 2 with coronal and sagittal acquisitions are presented in Figure 8. Severe image blurring (red arrows) is noticed at the diaphragm on the motion-averaged images, and the diaphragm becomes clearer and sharper on the MoraNet-FS, MoraNet-AF2 and MoraNet-AF4 images, which is in good agreement with the results of Figure 7.

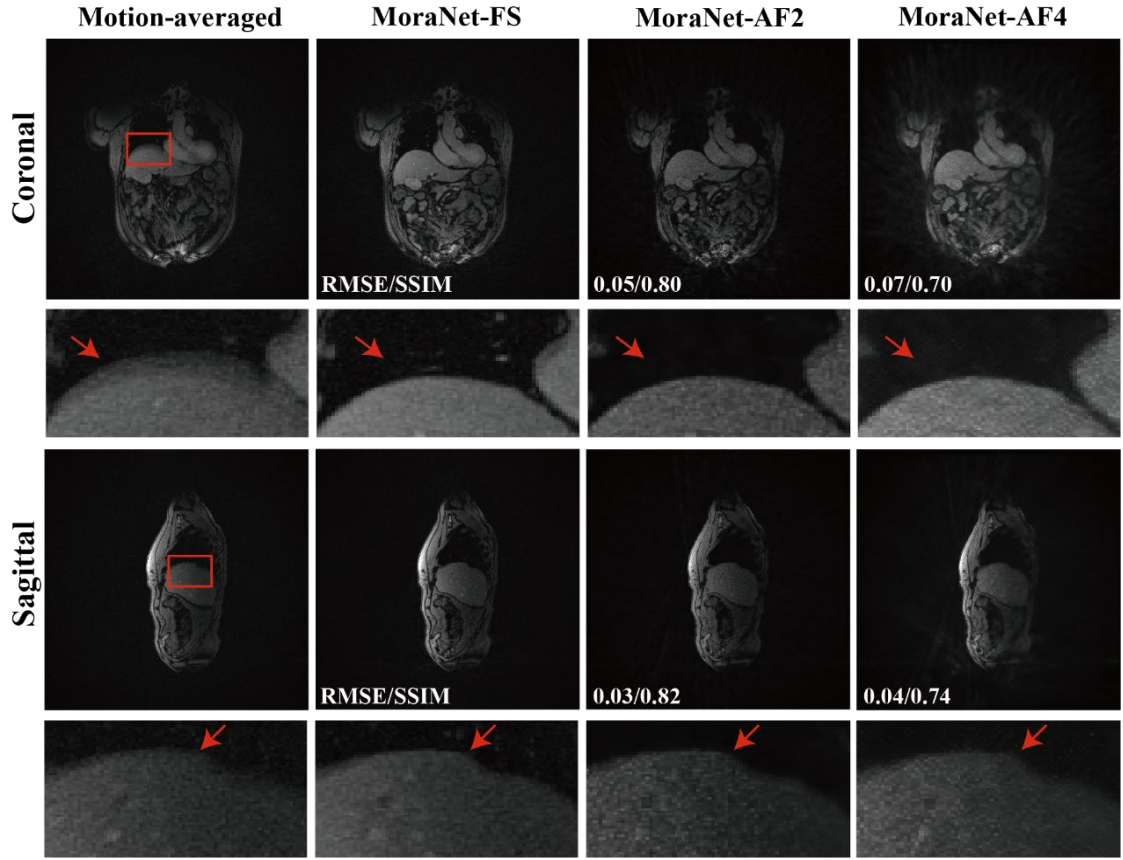

Figure 8 Volunteer chest image reconstructions on the fully sampled (motion-averaged and MoraNet-FS) and prospectively undersampled binned k-space data with AF = 2 (MoraNet-AF2) and AF = 4 (MoraNet-AF4) on coronal and sagittal planes. Zoomed regions (red rectangle) are shown at the bottom of each reconstructed image and red arrows point the diaphragm details.

## 3.5 Computational efficiency

The MoraNet and CS algorithms were both implemented on a desktop computer with an Intel Xeon central processing unit (CPU) of 16 GB RAM and 3.7 GHz. The latency of the CS reconstruction method on an image size of 256×256 was approximately 30 s, while the MoraNet took only 3 s, demonstrating an over ten-fold improvement in computational efficiency. In addition, the MoraNet was also executed on a high-performance computer equipped with an Nvidia Tesla V100 P32 GPU, and the inference time was around 300 ms, showing great potentials for routine clinical applications.

## 4. Discussion

Respiratory motion presents a major technical challenge, particularly in abdominal and pulmonary MR imaging. High-quality motion-resolved images will be essential for accurate clinical diagnoses and treatment planning [37]. Breath-holding acquisitions are normally performed in routine clinical practice to avoid respiratory motions [38,

39]. However, breath-holding scans limit the acquisition time and are typically infeasible for patients with comorbidities [40]. Respiratory gating is an alternative to minimize motion artifacts, which requires external respiratory bellows or MR navigators to track patients' respiratory motion and acquires data at a particular motion state (e.g., end-expiration) [41-43]. The respiratory gating method increases the total scan time and results in low imaging efficiency. Here, we developed an interpretable deep unrolled network to reconstruct motion-resolved images from successive radial acquisitions with free-breathing. 2D low-resolution dynamic images were firstly reconstructed in the proposed workflow to obtain the respiratory signal, and thus no additional external sensors or bellows are required to track patients' respiratory. Different spokes were used to reconstruct the low-resolution images and the accuracy of estimated motion signals were compared. Digital CoMBAT and experimental motion phantom results have shown that the spokes1 and spokes2 have better motion tracking ability than the spokes3. As the spokes2-reconstruction used only half spoke number of the spokes1-reconstruction, the temporal resolution of the motion tracking pipeline is 320 ms. In addition, the inference time is less than 0.5 s, which will be particularly beneficial for real-time image guidance of moving targets during radiotherapy treatments [44-46]. Prospective k-space data were acquired with tiny golden-angle (20.89°) and golden-angle (111.25°) radial trajectories on 1T and 1.5T MRI scanners, and high-resolution motion-resolved images were reconstructed at stage 2. The evaluations on multiple MRI scanners demonstrate the promises of the clinical deployment.

The proposed MoraNet was trained on single-channel MRI data (i.e., input single-channel k-space and output single-channel image) and tested on the multi-channel motion phantom and volunteer chest data. The k-space data from each channel was fed into MoraNet for image reconstruction, followed by a SoS operation on the channel-wise reconstructed images to generate the coil-sensitivity-combined image. Therefore, the coil sensitivity profile is not explicitly required in the network. Some studies include coil sensitivity information in the training process, however they often require model re-training for different coil configurations [47, 48]. The potential benefits of incorporating coil sensitivity information can be investigated in the future to further improve the network performance.

Due to the respiratory motion, the disease diagnoses of liver and pulmonary sites have always been an issue in clinical practice. Strain imaging (e.g., cine-tagging [49] and strain-encoded imaging [50]) has shown that the liver deformation during physiological motion can be used as a potential biomarker to stage liver fibrosis. Some research [51, 52] applies the elastic registration of inspiratory-to-expiratory lung MRI images to assess pulmonary function. However, the breath-hold acquisitions are required in these imaging techniques. The presented MoraNet pipeline enables dynamic imaging with free-breathing, and organ movement and deformation during respiratory circles can be visualized, which will typically facilitate the clinical diagnoses of liver and pulmonary diseases.

## 5. Conclusion

In this work, we developed and investigated a two-stage respiratory motion-resolved radial MR image reconstruction pipeline, built upon an interpretable deep unrolled network. Evaluations on the digital CoMBAT phantom, QUASAR motion phantom and volunteer chest data demonstrated that the proposed MoraNet enabled accurate motion signal estimation and effective motion artifacts reduction. Compared with the conventional CS-based method, the MoraNet could provide better image reconstruction performance and significantly shorten the computational time. The MoraNet shows great potential for improving abdominal and pulmonary imaging in clinical practice.

## Acknowledgement

This work was supported by the National Natural Science Foundation of China under Grant 62301352 and Grant 62301616. D.E.J.W. is supported by an Australian National Health and Medical Research Council Investigator Grant 2017140. P.K. is supported by an Australian National Health and Medical Research Council Investigator Grant 1194004.